\documentclass[aps, prl, twocolumn, superscriptaddress, showpacs]{revtex4}
\usepackage{amsmath}
\usepackage{amssymb}
\usepackage{graphicx}
\usepackage{fixmarks}

\begin{document}

\bibliographystyle{apsrev}

\title{﻿Jahn-Teller / Kondo Interplay in a Three-Terminal Quantum Dot}

\author{Ryan C. Toonen}
\email[email: ]{rctoonen@wisc.edu}
\affiliation{Laboratory for Molecular-Scale Engineering, Department of Electrical \& Computer Engineering, University of Wisconsin, Madison, Wisconsin 53706-1607}
\affiliation{The Wisconsin Solid State Quantum Computing Group; Departments of Physics, Electrical \& Computer Engineering, Materials Science \& Engineering, and Computer Science; University of Wisconsin, Madison, Wisconsin 53706-1390}
\author{Hua Qin}
\affiliation{Laboratory for Molecular-Scale Engineering, Department of Electrical \& Computer Engineering, University of Wisconsin, Madison, Wisconsin 53706-1607}
\author{Andreas K. H\"uttel}
\affiliation{Center for NanoScience, Department Physik, Ludwig-Maximilians-Universit\"at, 80539 M\"unchen, Deutschland}
\author{Srijit Goswami}
\affiliation{The Wisconsin Solid State Quantum Computing Group; Departments of Physics, Electrical \& Computer Engineering, Materials Science \& Engineering, and Computer Science; University of Wisconsin, Madison, Wisconsin 53706-1390}
\author{Daniel W. van der Weide}
\affiliation{The Wisconsin Solid State Quantum Computing Group; Departments of Physics, Electrical \& Computer Engineering, Materials Science \& Engineering, and Computer Science; University of Wisconsin, Madison, Wisconsin 53706-1390}
\author{Karl Eberl}
\affiliation{Max-Planck-Institut f\"ur Festerk\"orperforschung, 70569 Stuttgart, Deutschland}
\author{Robert H. Blick}
\affiliation{Laboratory for Molecular-Scale Engineering, Department of Electrical \& Computer Engineering, University of Wisconsin, Madison, Wisconsin 53706-1607}
\affiliation{The Wisconsin Solid State Quantum Computing Group; Departments of Physics, Electrical \& Computer Engineering, Materials Science \& Engineering, and Computer Science; University of Wisconsin, Madison, Wisconsin 53706-1390}

\begin{abstract}
In quantum dot circuits, screening electron clouds in strongly-coupled leads will hybridize with the states of the artificial atom.  Using a three-terminal geometry, we directly probe the atomic structure of a quantum dot with Kondo coupling.  Our measurements reveal that this hybrid system behaves as a non-linear entity with orbital degeneracy.  Geometric distortion reduces the symmetry, lowers the energy, and lifts the degeneracy---the Jahn-Teller effect.\end{abstract}


\maketitle

Recent measurements of vertical quantum dots (QDs) \cite{Sasa04, Sasa05} and carbon nanotube single-electron-transistors \cite{Jari05} have revealed Kondo transport phenomena due to orbital degeneracy arising from chiral symmetry.  Laterally gated QDs in ideal isotropic confinement potentials theoretically display a high degree of symmetry.  In practice however, the attachment of quantum point contacts (QPCs) reduces the spatial symmetry necessary for orbital degeneracy.  In the special case of a triangular configuration, degeneracies due to point group symmetry can be more easily achieved \cite{Zara03}.  Experiments have shown that Kondo co-tunneling allows for out-of-equilibrium transport through the excited energy levels of few electron QDs \cite{Sasa00, DeFr01, DeFr02, Koga03, Zumb04, Gran05}.  Theoretical predictions \cite{Sun01, Leba01, Sanc05, Shah05} and experimental investigations \cite{Letu05} have demonstrated the possibility of using a third lead to probe the density of states of a QD by means of cotunneling spectroscopy.  In this investigation, we use a three-lead configuration to explore the transport properties of a few-electron QD with Kondo interaction.  We adjust the Fermi level of the QD to create an artificial atom with four electrons.  As expected, exchange coupling yields a valence triplet state with lower energy than the singlet state.  When the QD lies in a configuration of equilateral symmetry with respect to the probing leads, the ground state is a hybridized doublet.  However, according to the Jahn-Teller theorem, partially-filled degenerate ground states in systems with nonlinear geometries are unpreferable.  Thus, we find that the symmetry of our system is spontaneously distorted in order for the degeneracy to split and stability to be achieved.  By applying a magnetic field perpendicular to the plane of the QD, we restore the ground state doublet.  To first order, our experimental data is in strong agreement with theory which has been developed for addressing the issue of Jahn-Teller / Kondo interplay from tunneling impurities with triangular symmetry \cite{Mous97}.

\begin{figure}
\includegraphics[width = 6.45cm]{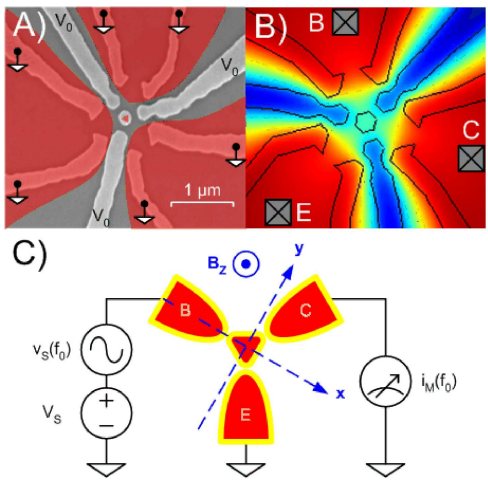}
\caption{\label{fig1} (A) SEM micrograph of the nanostructure.  An electrically floating center gate enhances the symmetry of the device by creating a uniform voltage potential in the vicinity of the QD.  (B) Illustration of the function of the coplanar shielding gates.  A voltage of $-1 V$ is applied to each constriction gate---fully depleting the QD.  This picture is based on a simplified FEMLAB 3.1 simulation which solved the Laplace equation in three dimensions for specified boundary conditions.  The color spectrum represents the full range of voltage potentials present in the simulated system (\textit{RED~$\to$~0} and \textit{BLUE~$\to$~-1~V}).  (C) Common-emitter circuit topology.  A small signal (with an amplitude of $5 {\mu}V_{RMS}$ at a frequency of $f_{0}=103 Hz$) is combined with a DC bias and applied to the base terminal.  The signal is detected at the collector with a current preamplifier and lock-in amplifier.}
\end{figure}

Our QD was formed by laterally constricting a two-dimensional electron gas (2DEG)---located $90$~$nm$ beneath the surface of a modulation-doped Al$_{x}$Ga$_{1-x}$As$/$GaAs heterostructure---with $36$-$nm$-high, Ni-Cr-Au-alloy Schottky split-gates (as shown in the SEM micrograph of Figure~\ref{fig1}(A)).  The results reported in this paper were measured in a \textit{He-3} one-shot cryostat.  Shubnikov-de-Haas and Quantum Hall measurements revealed that our heterostructure has a sheet density of $1.1$~$\times$~$10^{15}$~$m^{-2}$ and a mobility of $75$~$m^{2}$~$/$~$V-sec$ when the electron temperature is approximately $1.5$~$K$.

Our particular device has three unique features.  There is a floating gate positioned directly above the QD.  This metal island---having a diameter of $200$~$nm$---is capacitively coupled to the constriction gates; it \textit{flattens} the potential profile throughout the QD which effectively increases the symmetry of the system and creates a very shallow confinement well.  It has been previously demonstrated that shallow well QDs are ideal for confining a small number of electrons \cite{Cior02}.  As depicted in Figure~\ref{fig1}(B), the second unique feature of our device is the coplanar shielding of each active constriction gate by two electrically grounded gates.  This technique reduces collimation and allows for the 2DEG to be situated in close proximity to the QD---enabling a stronger interaction between the QD and lead states.  The third unique feature of our system is the fact that there is no single gate dedicated to \textit{plunging} the Fermi level of the QD.  Instead, two of the constriction gates are set to a constant value ($V_{0}$) and the third gate is varied around this value ($V_{GATE}=V_{0}~\pm~\Delta~V_{0}$).  It is true that this set-up does not allow us individual control over the strength of the tunneling barriers and the QD Fermi level.  However, by using fewer active gates, we were able to create a small enough three-port QD with a sufficiently large energy level spacing for resolving cotunneling spectral lines at temperatures on the order of $800~mK$.

As shown in Figure~\ref{fig1}(C), we operated our device in a manner similar to that of a bipolar junction transistor (BJT) in a common-source topology.  To form our QD, we first defined a base-port QPC with two of the constriction gates.  We varied the voltage on these gates until the conductance of the QPC channel was tuned to the lowest step of the QPC staircase.  We then applied a voltage ($V_{GATE}$) to the third constriction gate (situated between the collector and emitter ports).  We pre-conditioned the QD by changing this voltage from zero to a sufficiently large value ($-600 mV$) which fully depleted the system.  Finally, we lowered this gate voltage so that a conduction channel re-opened between the collector and emitter leads, and a few electrons could occupy the QD. 

Experiments performed by Simmel, \textit{et al.} have shown that the energy levels of a QD will pin with respect to that of a strongly coupled lead \cite{Simm99}.  In our case, the energy levels of the QD appear to have been pinned with respect to the collector and emitter leads.  We measured the differential conductance, $G={i_{M}/v_{S}}~{\approx}~{dI_{M}/dV_{S}}$
, across the base and collector as a function of the DC voltage bias on the base lead.  When this chemical potential was in resonance with excited energy levels of the QD, electrons were injected from the base into the QD thereby enabling us to directly probe the QD density of states.

\begin{figure}
\includegraphics[width = 6.45cm]{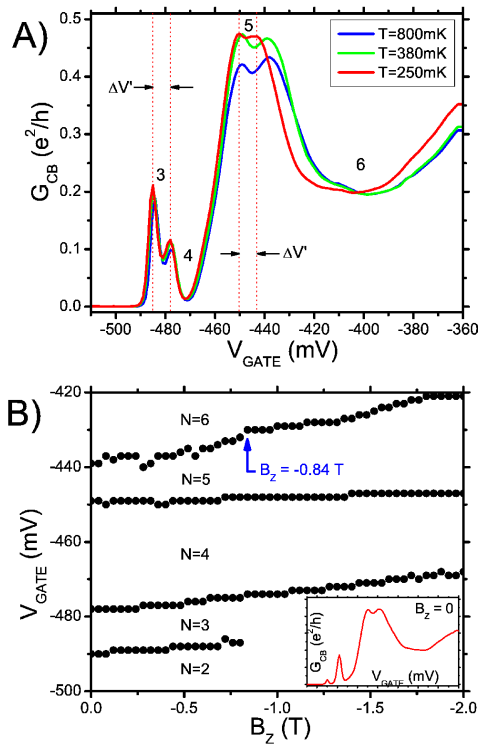}
\caption{\label{fig2} Evidence for the Kondo co-tunneling.  (A) Kondo spectra with zero bias across the QD.  The gate-controlled spectrum is measured at three different temperatures.  A decrease in temperature results in an increase of quantum coherence and an increase of the Kondo-conduction feature.  At the lowest electron temperature we achieved ($T~=~250~mK$), the $N~=~3$ and $N~=~5$ Coulomb valleys both had widths of approximately ${\Delta}V_{GATE}~=~{\Delta}V'~=~7~mV$.  These three traces were taken within a few hours of each other as we raised the base temperature of the cryostat.  (B) Fock-Darwin (base-to-collector) conductance spectra.  A kink in the peak trace \textit{above} the $N=5$ Coulomb valley suggests a change in filling factor at $B_{Z}~=~-0.84~T$.  The general lack of kinks in the spectra provides evidence that the QD is biased in the few-electron limit.}
\end{figure}

Figure~\ref{fig2}(A) displays zero-bias measurements of the base-to-collector conductance for three different temperatures.  Decreasing the temperature of the system increased the quantum coherence between lead states and QD states.  As a result, we observed an enhancement of cotunneling conductance within the Coulomb valleys containing an odd number of electrons\cite{Gold98, Cron98}.  Figure~\ref{fig2}(B) shows the recorded peak traces for this zero-bias measurment as a function of a magnetic field (the Fock-Darwin spectra).  The general lack of kinks suggests that our QD is in the few-electron regime \cite{Kouw01}.

\begin{figure}
\includegraphics[width = 8.6cm]{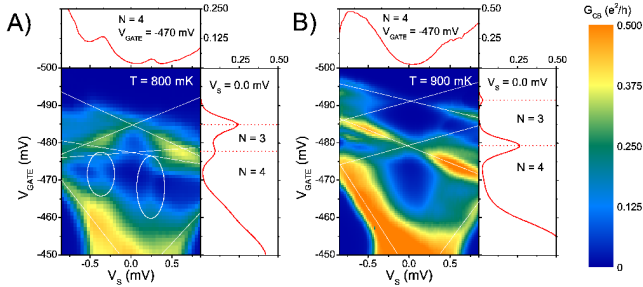}
\caption{\label{fig3} Coulomb diamonds outlined with dotted lines.  (A) Contour plot of Coulomb diamond with Kondo cotunneling features (circled) at $T~\sim~800~mK$.  (B) A plot taken with the same biasing conditions as (A) but with a higher temperature ($T~\sim~900~mK$).  The Kondo signatures are obscurred by the background conductance due to inelastic cotunneling.}
\end{figure}

To investigate the out-of-equilibrium behavior of our device, we measured the conductance as a function of both gate voltage and bias applied to the base node of our device.  Figures~\ref{fig3}(A) and (B) show contour plots of Coulomb diamonds taken with the same biasing conditions but at two different temperatures.  In Figure~\ref{fig3}(A), there are prominent co-tunneling spectral lines within the diamond containing the slice $V_{GATE}~=~-470~mV$.  As shown in Figure~\ref{fig3}(B), these distinctive lines fade into the background conductance caused by inelastic cotunneling as the electron temperature of the 2DEG is increased.  We also see from these plots that the small diamond centered on $V_{GATE}~=~-480~mV$ in Figure~\ref{fig3}(A) vastly expands as a function of temperature (Figure~\ref{fig3}(B)).  The observed temperature dependent co-tunneling suggests that our system demonstrated Kondo coupling between the QD and electron clouds in strongly coupled leads.  

We noticed that there was a rather large Coulomb valley centered about the slice $V_{GATE}~=~-400~mV$.  When both the \textit{s}- and \textit{p}-shells are completely filled \cite{Zhu96}, a QD will contain a \textit{magic number} of six electrons and the increased amount of energy required for ionization will yield a Coulomb valley larger than its nearest and next-nearest neighbors.  We therefore formed the conjecture that this valley corresponds to the case of a six-electron QD system as a means of justification for fitting our cotunneling data to a four-electron model.  We did not resolve the boundaries of the one- and two-electron QD because in these cases the spectral lines had magnitudes below our threshold of detection.

We were particularly interested in studying the out-of-equilibrium cotunneling peaks present in the four-electron valley of Figure~\ref{fig3}(A).  We tracked the centers of these spectra lines as a function of magnetic field applied perpendicular to the plane of the QD (Figure~\ref{fig3}).  We plotted only the points which were clearly distinguishable from the background conductance.  The application of a magnetic field $B_{Z}$ perpendicular to the plane of the QD allowed us to control the splitting of the doublet state.  Around the point of restored degeneracy, $B_{0}$, we observed a crossing of the doublet states and an anti-crossing of the excited singlet and triplet states.  Since these peaks lied within the four-electron Coloumb diamond, we fit their traces to the Jahn-Teller-Kondo model of Moustakas and Fisher \cite{Mous97} for describing the energy spectra of a four-electron tunneling impurity with triangular symmetry.  We adopted their Hamiltonian $H_{center}$ for describing the system in a state of equilateral symmetry, and we allowed the constants of their splitting term $H_{split}$ to have linear dependence on the perpendicularly applied magnetic field.

\begin{figure}
\includegraphics[width = 8.6cm]{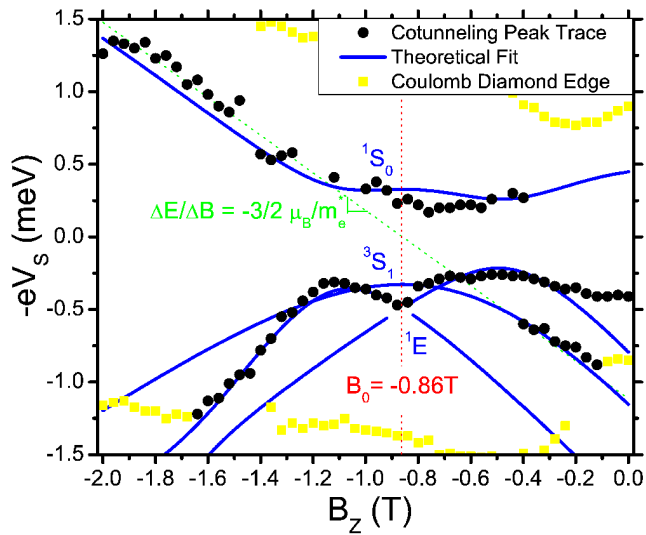}
\caption{\label{fig4} Experimental co-tunneling conductance data.  Peak traces of co-tunneling (base-to-collector) spectral lines taken in the four-electron QD valley.  We display the theory (blue lines) versus experiment (black dots).  The yellow squares mark ionizing transitions and define the edge of the Coulomb diamond.}
\end{figure}

The Moustakas-Fisher model describes distortion of a system with $C_{3V}$ point-group symmetry to one with $C_{2V}$ symmetry.  The disortion can occur along any one of the three axes of an equilateral triangle.  This possibility gives rise to three single-particle states: \textit{S} which has a symmetric distribution about the three states of distortion; and \textit{L} and \textit{R} which respectively traverse the states of distortion with left- and right-handed polarity.  A constant $\epsilon$ lowers the energy of \textit{S} with respect to \textit{L} and \textit{R} which are degenerate when no magnetic field is applied.  A ferromagnetic exchange constant $J$ accounts for Hund's first rule by describing the tendency for valence spins to align within the quantum dot---thus ensuring that the triplet state will have a lower energy than the singlet.  Likewise, a constant $K$ accounts for exchange coupling with the \textit{S} state.  Effectively, it creates a hybridized doublet which becomes the ground state when the condition $2K^{2}~>~{\epsilon}{J}$ is met.  Distortion of the impurity away from equilateral symmetry is described by the mixing of single-particle states with energies $\delta_1$ (for \textit{L} and \textit{R}) and $\delta_2$ (for \textit{S} and \textit{R} or \textit{S} and \textit{L}) which are linear in spatial displacement.  In this analysis, only the four lowest energy states of the composite system are of concern.  As represented in Figure~\ref{fig4}, they are the valence spin triplet \textit{${}^{3}S_{1}$}, the valence spin singlet \textit{${}^{1}S_{0}$}, and the \textit{${}^{1}E$} hybridized doublet pair.  The energies associated with the \textit{${}^{3}S_{1}$}, \textit{${}^{1}S_{0}$}, and \textit{${}^{1}E$} states are $-(2\epsilon+J)$, $-(2\epsilon-J)$, and $-\epsilon~-~(\epsilon+J)/2~-~\sqrt{({\epsilon}-{J})^{2}/2^{2}~+~2K^{2}}$.

To fit our data, we applied the Moustakas-Fisher model \cite{Mous97} with a Cartesian offset of ($B_{0}=-0.86~T$, $E_{0}=2\epsilon$).  By inspection, we found that our measurments are in strong agreement with this theory around the point of restored doublet degeneracy ($B=B_{0}$) when the physical parameters of $H_{center}$ are set as follows: ${\epsilon}=2.62~eV$, $J={\epsilon}/8$, and $K={\epsilon}/3$.  To account for the influence of an applied magnetic field on the geometric distortion of our system, we replaced the coefficients of $H_{split}$ with magnetic-field-dependent expressions, ${\delta_{1,2}}{\rightarrow}{g_{1,2}}{\cdot}{(e/{\sqrt{8{m_e}})}}{\cdot}{B}$, and we expressed the position variables in terms of the QD energy, ${x}{\pm}{iy}{\rightarrow}{r{\cdot}(cos({\pi}/4) {\mp}{i}sin({\pi}/4))}$ where ${r^{2}}={\hbar}/({m_{e}^{\ast}}{\sqrt{{\omega^2}+{\omega_{C}^{2}}}})$ \cite{Kouw01}, $m_{e}^{\ast}$ is the effective mass constant of electrons in GaAs, and $\omega_{C}$ is the cyclotron frequency.  We empirically determined $g_{1}$ and $g_{2}$ to be $5/2$ and $7/2$, respectively.  An estimate of the energy associated with Jahn-Teller distortion of our system can be determined from the Cartesian offset of our data, $E_{JT} \approx {\frac{3}{2}}{\frac{{\mu}_{B}}{m_{e}^{\ast}}}{B_{0}} \approx 1.1 meV$.

Kondo cotunneling in a three-terminal configuration is the mechanism which has enabled us to study the spectra of excited energy levels for a four-electron QD.  From an engineering point of view, we have demonstrated that the influence of Jahn-Teller distortion is a factor which must be considered when designing QD circuits for practical applications.  We have also shown that we can tune our QD to specific energy states by means of adjusting the base and gate voltages of our device.  Our experiments indicate the viability of controlling spin-entanglement properties of electrons ejected from our device.  Extensive theory has been developed concerning the topic of correlated tunneling events in three-terminal QD systems with ferromagnetic leads \cite{Cott04a, Cott04b}.  Additionally, theorists have conceptualized three-terminal QD network schemes for generating spatially separated pairs of entangled electron spins \cite{Cost01, Oliv02, Sara03, Zhan04}.

We thank Dr. Marta Prada and Prof. Robert Joynt for their insights.  Likewise, we extend our gratitude to the members of the Laboratory for Molecular-Scale Engineering and the Wisconsin Solid State Quantum Computing Group for providing us with thought-provoking discussions.  Funding for the project was provided in part by the National Science Foundation (DMR-0325634).

\bibliography{bibdata}

\end{document}